\documentclass[review]{elsarticle}

\usepackage{lineno,hyperref}
\modulolinenumbers[5]

\usepackage{multirow}
\usepackage{caption}
\usepackage{booktabs}
\usepackage{booktabs}
\usepackage{siunitx}

\usepackage{float}
\usepackage{longtable}

\usepackage{amsfonts}
\usepackage{amsmath}
\usepackage{xfrac}

\usepackage{graphicx}

\usepackage{algorithm}
\usepackage{algpseudocode}

\DeclareMathOperator{\dist}{dist}
\DeclareMathOperator{\ceil}{ceil}

\journal{Computer Physics Communication}

\begin{document}

\begin{frontmatter}

\title{Voronoi Particle Merging Algorithm for PIC Codes}

\author{Phuc T. Luu}
\ead{luu@tp1.uni-duesseldorf.de}
\author{T. T\"uckmantel}
\author{A. Pukhov}
\address{Theoretische Physik I, Heinrich Heine Universit\"at, 40225 D\"usseldorf, Germany}

\begin{abstract}
We present a new particle-merging algorithm for the particle-in-cell method. Based on the concept of the Voronoi diagram, the algorithm partitions the phase space into smaller subsets, which consist of only particles that are in close proximity in the phase space to each other. We show the performance of our algorithm in the case of the two-stream instability and the magnetic shower.
\end{abstract}


\end{frontmatter}

\linenumbers

\section{Introduction} \label{s_introduction}
For more than $60$ years the particle-in-cell (PIC) technique \cite{dawson_1983} has been used to simulate a wide variety of physical problems, ranging from electrical discharge to particle acceleration. However, in several scenarios - in particular field ionisation or QED cascades - the number of particles in the simulation box grows exponentially. Due to an overwhelming number of particles, the associated memory required can easily exceed that available on even high performance computers and as a consequence the computational performance drops drastically.

In these situations, a particle merging algorithm (PMA) has to be implemented. The main goal of a PMA is to reduce the number of particles in a simulation box while keeping the physical properties of the system intact after a merging event. A straightforward PMA is to randomly pick a pair of particles and then merge, see for example \cite{timokhin_merging_2010}. Since it merges with no guidance, the method is not able to preserve the phase space distribution, and so the physical picture is likely to be distorted after merging. The problem is that it fails to incorporate the notion of proximity in the phase space, i.e. how similar particles are, into its framework. In the scope of this paper, we call this PMA the blind method.

Lapenta already proposed a scheme for merging particles (called ``particle coalescence'') in \cite{lapenta_1994} and \cite{lapenta_2002}. In this method, particles are first sorted into two bins. Then the binning process continues until the number of particles per bin is small enough for the pairwise comparison. This type of PMA was then refined and improved by Teunissen and Ebert \cite{teunissen_merging_2014}, in which the k-d tree method was employed to search for the nearest neighbour. Recently, a similar approach was also proposed by Vranic et al. \cite{vranic_2015}, where the momentum space is divided into smaller subcells for sorting particles.

We design our PMA from a different point of view, in which the algorithm not only merges particles which are close in the phase space but also offers users a direct control over errors introduced by a merging event. The notion of proximity in our algorithm is developed through the concept of the Voronoi diagram \cite{voronoi_1908}, thus the name Voronoi PMA. As shown later, the quantification of the error is realised through the coefficients of variation. The algorithm is successfully implemented into the framework of the VLPL (Virtual Laser Plasma Laboratory) code \cite{pukhov_1999}.

The paper is organised as follows: in section \ref{s_voronoi}, we briefly introduce the definition and some examples of the Voronoi diagram; the comprehensive description of our PMA is revealed in section \ref{s_algorithm}; in section \ref{s_simulation} we test the performance of our merging algorithm with three cases: the counter-propagating plasma blocks, the two-stream instability, and the magnetic shower simulations; finally, we summarise the paper in section \ref{s_summary}.

\section{Voronoi diagram} \label{s_voronoi}

For any given set of $n$ sites, $S = \{s_1, s_2, ..., s_n\}$ in the real $d$-space $\mathbb{R}^d$, the Voronoi cell $\mathcal{V}_k$ associated with the site $s_k$ is a set of points in $\mathbb{R}^d$, such that the distance from those points to $s_k$ is not greater than the distance to any other site $s_j$ $(j \neq k)$ in $S$ \cite{dwyer_1988}.
\begin{align}
\mathcal{V}_k = \{ x \in \mathbb{R}^d \; | \; \forall j: \text{dist}(x, s_k) \leq \text{dist}(x, s_j) \} \; \text{for } 1  \leq i, j \leq n.
\label{eq:voronoi_def}
\end{align}
Here, $\text{dist}(x, y)$ denotes the metric function of the distance in $\mathbb{R}^d$. The Voronoi diagram was first developed, though informally, in 1644 by Descartes. In 1908, the Russian-Ukrainian mathematician G. F. Voronoi formally defined and studied the general case \cite{voronoi_1908}. The concept is used in many contemporary research fields, such as geophysics, meteorology, and condensed matter physics.
\begin{figure}[ht!]
	\includegraphics[width=100mm]{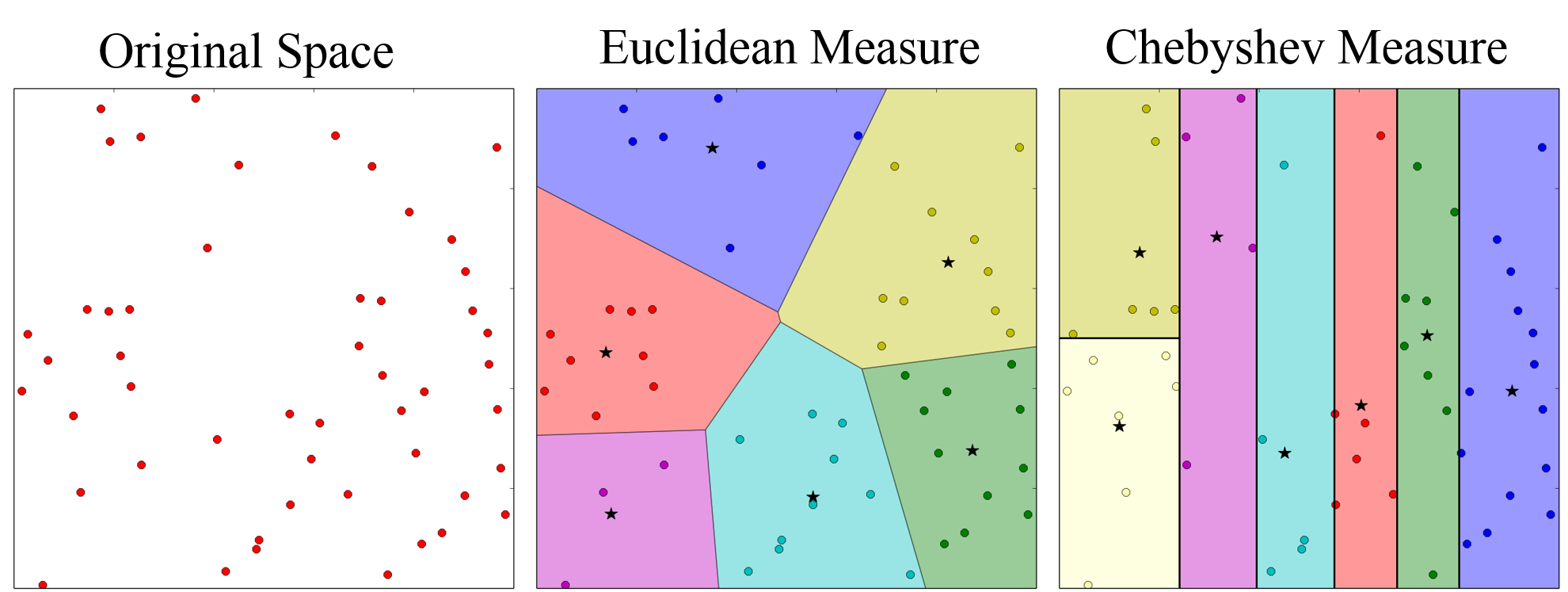}
	\centering
	\caption{The Voronoi diagrams with different metric functions. Each Voronoi region is painted with a distinct colour. The black star in each region is the Voronoi centroid.}
	\label{fig:dist_measure}
\end{figure}

Observing eq. \eqref{eq:voronoi_def}, we see that the metric function $\text{dist}(x, y)$ plays a vital role in the formation of the Voronoi diagram. Different metrics will result in different Voronoi diagrams. Moreover, in our case, different metrics also require different implementations of the algorithm (see section \ref{s_algorithm} for more detail). Fig. \ref{fig:dist_measure} shows the Voronoi diagram of a random distribution with Euclidean and Chebyshev measures. Given two vectors $\mathbf{p}$ and $\mathbf{q}$, the Euclidean distance is
\begin{align}
\dist(\mathbf{p}, \mathbf{q}) = \sqrt{ \sum_{i} (q_i - p_i)^2},
\end{align}
while the Chebyshev distance is given by
\begin{align}
\dist(\mathbf{p}, \mathbf{q}) = \max {|q_i - p_i|}. \label{eq:cheb_dist}
\end{align}

\section{Algorithm} \label{s_algorithm}
\textbf{Input:} The algorithm requires two user inputs, $T_\mathbf{X}$ and $T_\mathbf{P}$, which are the tolerances for position and momentum. These parameters are employed as the stopping condition and appear at step 3. A merging event will take place in a simulation cell if the particle number $N$ of that cell is greater than the minimum particle number $N_{\text{min}}$.

\textbf{Step 1:} For every simulation cell, collect all particles (weight $w_i$, position $\mathbf{x}_i$, and momentum $\mathbf{p}_i$) in that cell into a set $\mathcal{V}_0$. This set $\mathcal{V}_0$ is the first Voronoi cell, which covers the entire phase space of a simulation cell. We then calculate the statistical average in the phase space of this set of particles $\mathcal{V}_0$ by the following formulae:
\begin{align}
W_0 &= \sum_{i \in \mathcal{V}_0} w_i, \\
\mathbf{X}_0 &= \frac{\sum_{i \in \mathcal{V}_0}w_i \mathbf{x}_i}{\sum_{i \in \mathcal{V}_0}w_i},\\
\mathbf{P}_0 &= \frac{\sum_{i \in \mathcal{V}_0}w_i \mathbf{p}_i}{\sum_{i \in \mathcal{V}_0}w_i}.
\end{align}
The point $(\mathbf{X}_0, \mathbf{P}_0)$ with weight $W_0$ is the centroid of the first Voronoi cell $\mathcal{V}_0$. From now on, quantities of a Voronoi centroid are denoted by the capital letters.

\textbf{Step 2:} We calculate the standard deviation of each dimension $l$ in the phase space with respect to the current Voronoi centroid:
\begin{align}
\sigma_{\mathbf{X}_{0,l}} &= \sqrt{\frac{1}{W_0} \sum_{i} w_i ( \mathbf{x}_{i,l} - \mathbf{X}_{0,l} )^2 }, \\
\sigma_{\mathbf{P}_{0,l}} &= \sqrt{\frac{1}{W_0} \sum_{i} w_i ( \mathbf{p}_{i,l} - \mathbf{P}_{0,l} )^2 }.
\end{align}
We compute the coefficient of variation (CV) $\Delta$ for each dimension. The CVs for spatial and momentum dimensions are defined as
\begin{align}
\Delta_{\mathbf{X}_{0,l}} &= \frac{\sigma_{\mathbf{X}_{0,l}}}{L_{\mathbf{X}_{0,l}}}, \label{eq:cv_x}\\
\Delta_{\mathbf{P}_{0,l}} &= \frac{\sigma_{\mathbf{P}_{0,l}}}{\mathbf{P}_{0,l}}. \label{eq:cv_p}
\end{align}
For the spatial dimensions, due to the symmetry in space the CV $\Delta_\mathbf{X_0}$ is defined as the ratio between the standard deviation and the length $L_{\mathbf{X}0}$ of the first Voronoi cell $\mathcal{V}_0$. On the other hand, since there is no such symmetry in the momentum space, the CV $\Delta_\mathbf{P_0}$ is obtained from dividing the standard deviation by the mean value. As the CVs are dimensionless numbers we can treat the data obtained from the position and momentum spaces equally (see step 4 below). In our algorithm, the CVs represent the accuracy of the merging scheme, with smaller CVs resulting in smaller errors due to merging.

\textbf{Step 3:} We compare the recently obtained CVs $\Delta_{\mathbf{X}_0}$ and $\Delta_{\mathbf{P}_0}$ with their corresponding tolerances $T_\mathbf{X}$ and $T_\mathbf{P}$. If a Voronoi cell has all six CVs less than or equal to the tolerances, the algorithm will mark that cell finished and stop dividing it. On the other hand, as long as there is at least one component whose CV does not satisfy the aforementioned requirement, the algorithm will keep going to the next step.

\textbf{Step 4:} We consider the individual components of $\Delta_{\mathbf{X}_0}$ and $\Delta_{\mathbf{P}_0}$, that is $\{\Delta_x, \Delta_y, \Delta_z, \Delta_{p_x}, \Delta_{p_y}, \Delta_{p_z}\}$, and find the axis $k$ which has the largest deviation.
\begin{align}
k &= \max_{l} \Delta_l, \text{ with } l \in \{ x, y, z, p_x, p_y, p_z \}. \label{eq:step_4}
\end{align}

\textbf{Step 5:} Make a hyperplane cut through the  the Voronoi centroid perpendicular to the axis $k$. Denote $q$ and $Q$ the dynamic variables of the particles and of the centre, respectively, on the axis $k$. The hyperplane cut divides the set $\mathcal{V}_0$ into two new independent subsets $\mathcal{V}_1$ and $\mathcal{V}_2$, whose new centroids are given by
\begin{center}
\def\arraystretch{1.8}\arraycolsep=1.4pt
$\begin{array}{rl!{\quad\vline\quad}rl}
\mathcal{V}_1 &= \left \{i \in \mathcal{V}_0: q_i \leq Q \right\} &   \mathcal{V}_2 &= \left \{i \in \mathcal{V}_0: q_i > Q \right\} \\
W_1 &= \sum_{i \in \mathcal{V}_1} w_i &   W_2 &= \sum_{i \in \mathcal{V}_2} w_i \\
\mathbf{X}_1 &= \frac{\sum_{i \in \mathcal{V}_1} w_i \mathbf{x}_i}{\sum_{i \in \mathcal{V}_1} w_i} &   \mathbf{X}_2 &= \frac{\sum_{i \in \mathcal{V}_2} w_i \mathbf{x}_i}{\sum_{i \in \mathcal{V}_2} w_i} \\
\mathbf{P}_1 &= \frac{\sum_{i \in \mathcal{V}_1} w_i \mathbf{p}_i}{\sum_{i \in \mathcal{V}_1} w_i} & \mathbf{P}_2 &= \frac{\sum_{i \in \mathcal{V}_2} w_i \mathbf{p}_i}{\sum_{i \in \mathcal{V}_2} w_i}
\end{array}$
\end{center}

\textbf{Step 6:} Sort the particles into their corresponding new sets. Repeat steps 2-6 for the new sets $\mathcal{V}_1$ and $\mathcal{V}_2$ until the stopping condition is satisfied.

\textbf{Step 7:} If the stopping condition is met for all Voronoi cells, the algorithm removes all particles from the simulation cell and replaces them with the Voronoi centroids as the merged particles. The algorithm ends here.

We have several remarks on our algorithm:
\begin{itemize}
\item Our Voronoi PMA is inspired by Schreiber's adaptive k-means clustering algorithm used in Computational Geometry \cite{schreiber_adaptive_voronoi}.
\item In step 1, we state that the merging process is carried out cell by cell. However, the algorithm can be adjusted such that the first Voronoi cell $\mathcal{V}_0$ contains all particles from the simulation box and starts merging from there. The rest of the algorithm is kept intact. However, it is likely that the global merging approach violates the local charge conservation. In this case, one must take into account a correction scheme in order to compensate for the error caused by merging events. Which implementation is used depends strongly on the user preference or the code framework. We adhere to the cell-by-cell implementation as it is readily parallelised.
\item The distance measure used here (see eq. \eqref{eq:step_4}, step 4) can be considered as a Chebyshev-like distance, since eq. \ref{eq:cheb_dist} is not guaranteed for every particle and phase space dimension. We have chosen this measure instead of a more obvious candidate, the Euclidean measure, for the following reasons:
\begin{enumerate}
\item The simplest implementation of the Euclidean measure requires the seeding of Voronoi centroids at the beginning of the algorithm. Moreover, the number of Voronoi centroids is kept constant throughout the algorithm. This limitation not only reduces greatly the flexibility of the algorithm but also cannot fit well to the dynamic situation of a physical problem \cite{schreiber_adaptive_voronoi}. Conversely, the Chebyshev measure requires no seeding and suits perfectly the divide-and-sort scheme, which is applied here.
\item In \cite{mardia_multivariate_analysis}, the author states a rule of thumb that for a given dataset of $N$ points, the number of centroids is set to $ k \approx \sqrt{N/2}$. Again, the number of Voronoi centroids cannot be changed once the algorithm starts. As such, we do not follow this rule.
\item In order to use the Euclidean measure without a fixed number of centroids, we would have to solve the problem of an unknown number of clusters in a dataset. This can be done through the Bayesian information criterion \cite{pelleg_2000} or the removing centroids method \cite{bischof_1999}. The former approach is difficult to implement, while the latter tends to be computationally intensive.
\end{enumerate}
\item In the momentum space, the Voronoi PMA groups particles by taking into account both the direction and the magnitude of particles' momenta. Due to the difference in the direction, it might occur that the energy is lost after a merging event. The relative error in the total energy is observed in Fig. \ref{fig:tsi_energy} for the two-stream instability and Fig. \ref{fig:voronoi_compare_w_legend} for the magnetic shower below. These graphs show that the loss in energy per merging event is extremely small. However, the merging quality can be further improved by introducing a mechanism to conserve energy perfectly and directly. One can consider the Langdon-Marder corrector-scheme \cite{langdon_1992}\cite{wang_2009}\cite{wang_2010} or follow the proposal to merge into two particles \cite{vranic_2015}. We also make a side remark that the Langdon-Marder scheme becomes obligatory in case users want to implement the algorithm through the global merging approach.
\end{itemize}
\section{Simulation} \label{s_simulation}
Having presented the algorithm, we proceed to test its performance. To this end, we consider three situations: counter-propagating plasma blocks, the two-stream instability \cite{bittencourt_2004} \cite{birdsall_2004}, and the magnetic shower produced by an energetic particle entering a strong magnetic field \cite{anguelov_showers_1999}.

Before going further, we briefly describe the implementation of the blind method used here for comparison. We define the parameter $\alpha$ as the merging fraction. A merging event will take place in a simulation cell if the number of particles $N$ of that cell satisfies the condition $N > \ceil(\alpha N)$. Then, the blind method merges particles in the current cell until the number of particles after merging is at maximum $\ceil(\alpha N)$. This implementation allows the blind method produces the same number of particles as in the Voronoi PMA for fair comparison.
\subsection{Counter-propagating Plasma Blocks}
\begin{figure}[ht!]
	\includegraphics[width=100mm]{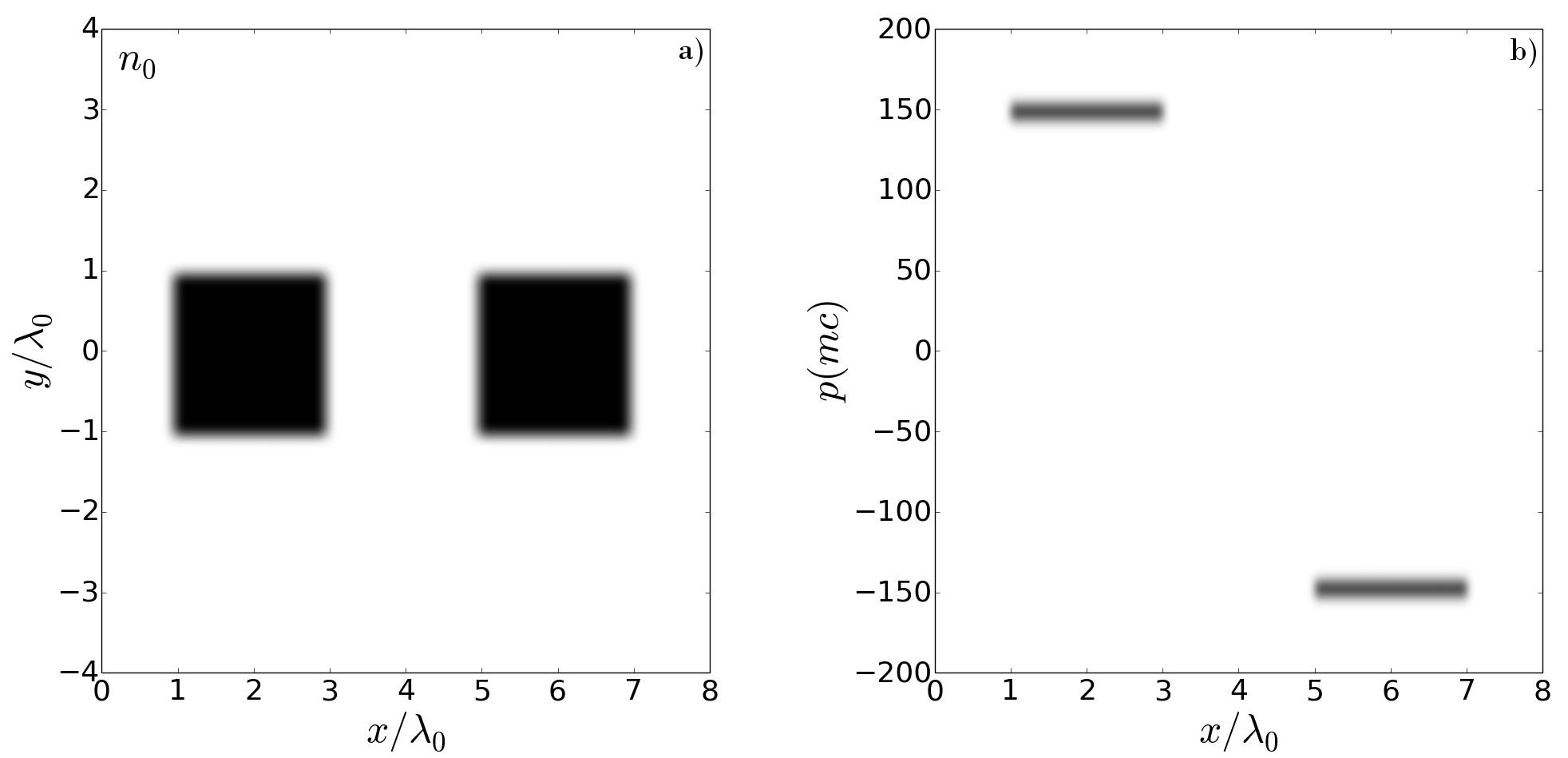}
	\centering
	\caption{The initial configuration for the counter-propagating plasma blocks simulation: two blocks have the same momentum magnitude but opposite propagation directions. The merging event will commence when two blocks start overlapping each other since the particle number exceeds the threshold. For the Voronoi PMA, the threshold is $N_{min} = 15$, and $\ceil(\alpha N)$ for the blind method. A good merging algorithm will leave behind no change in the phase space distribution apart from the translation in the $x$-direction.}
	\label{fig:block_dist_first_frame}
\end{figure}
\begin{figure}[ht!]
	\includegraphics[width=100mm]{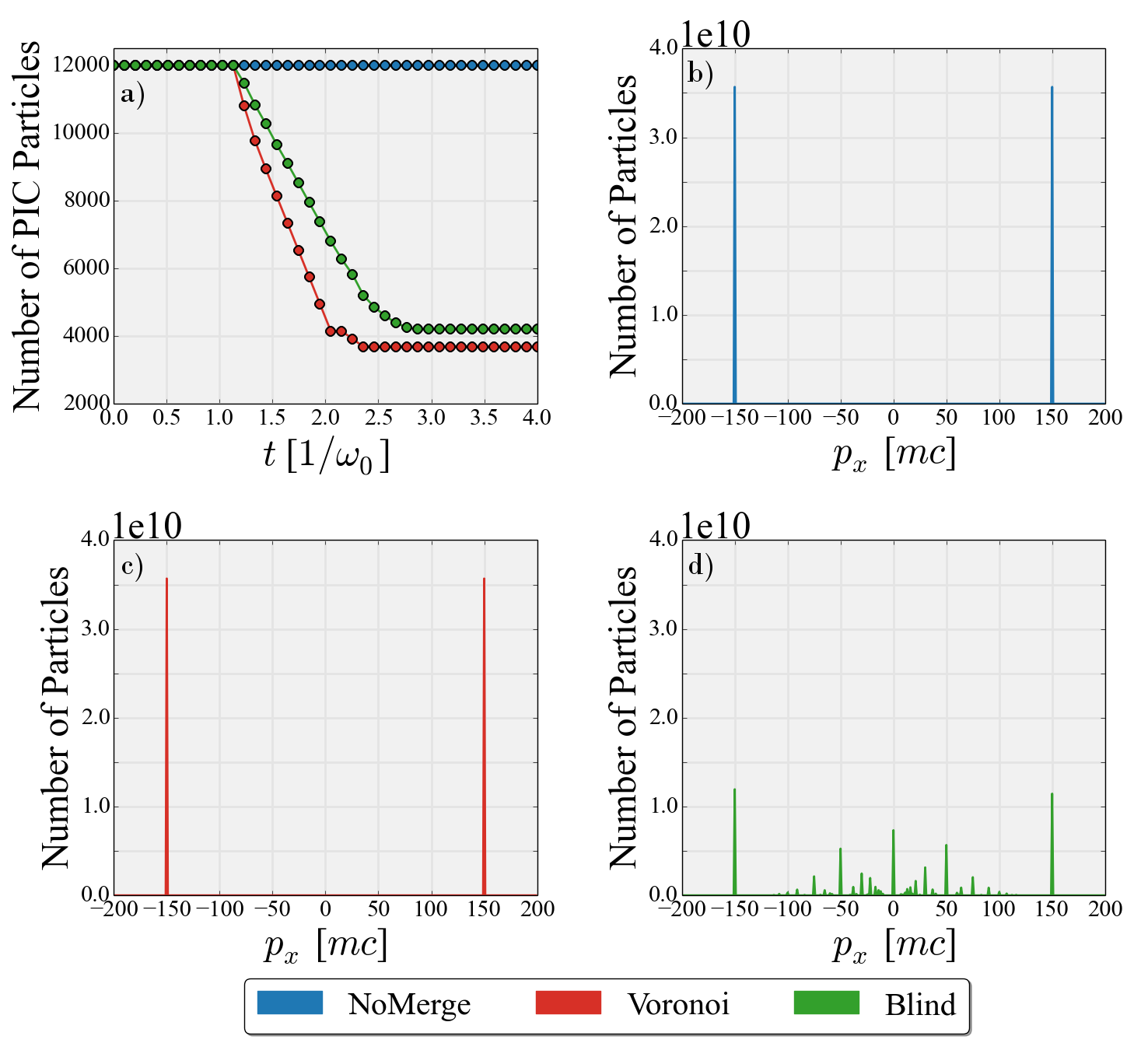}
	\centering
	\caption{The number of PIC particles during the simulation (fig. a) and the histograms for no-merge, the Voronoi PMA, and the blind method (figs. b, c, and d respectively) in the counter-propagating plasma blocks simulation. Despite merging into a similar number of particles, the Voronoi PMA does not distort the momentum distribution.}
	\label{fig:block_N_spectrum_w_legend}
\end{figure}
The counter-propagating plasma blocks simulation is a simple test, in which two blocks of non-interacting particles with uniform density distribution propagate and then overlap each other. These blocks have the same momentum magnitude but opposite propagation directions (see Fig. \ref{fig:block_dist_first_frame}). With no merging, there is no change to the system apart from the translation in $x$-direction after the blocks pass through each other. By using this test we can easily spot whether a given PMA preserves the phase space distributions since there is a duration when the blocks overlap. If a merging method does not preserve, two or more particles from the different distributions might be merged together. Here, we compare the performance of the Voronoi PMA and the blind method. The merging period $T_{mrg} = 2 \Delta t$, with $\Delta t$ is the time step, is applied for both methods. For the Voronoi PMA, the tolerances are $T_\mathbf{X} = 0.4$ and $T_\mathbf{P} = 0.01$. For the blind method, we deliberately choose the parameter $\alpha$ so as to give a similar final number of particles as in the Voronoi PMA.

\begin{figure}[ht!]
	\includegraphics[width=120mm]{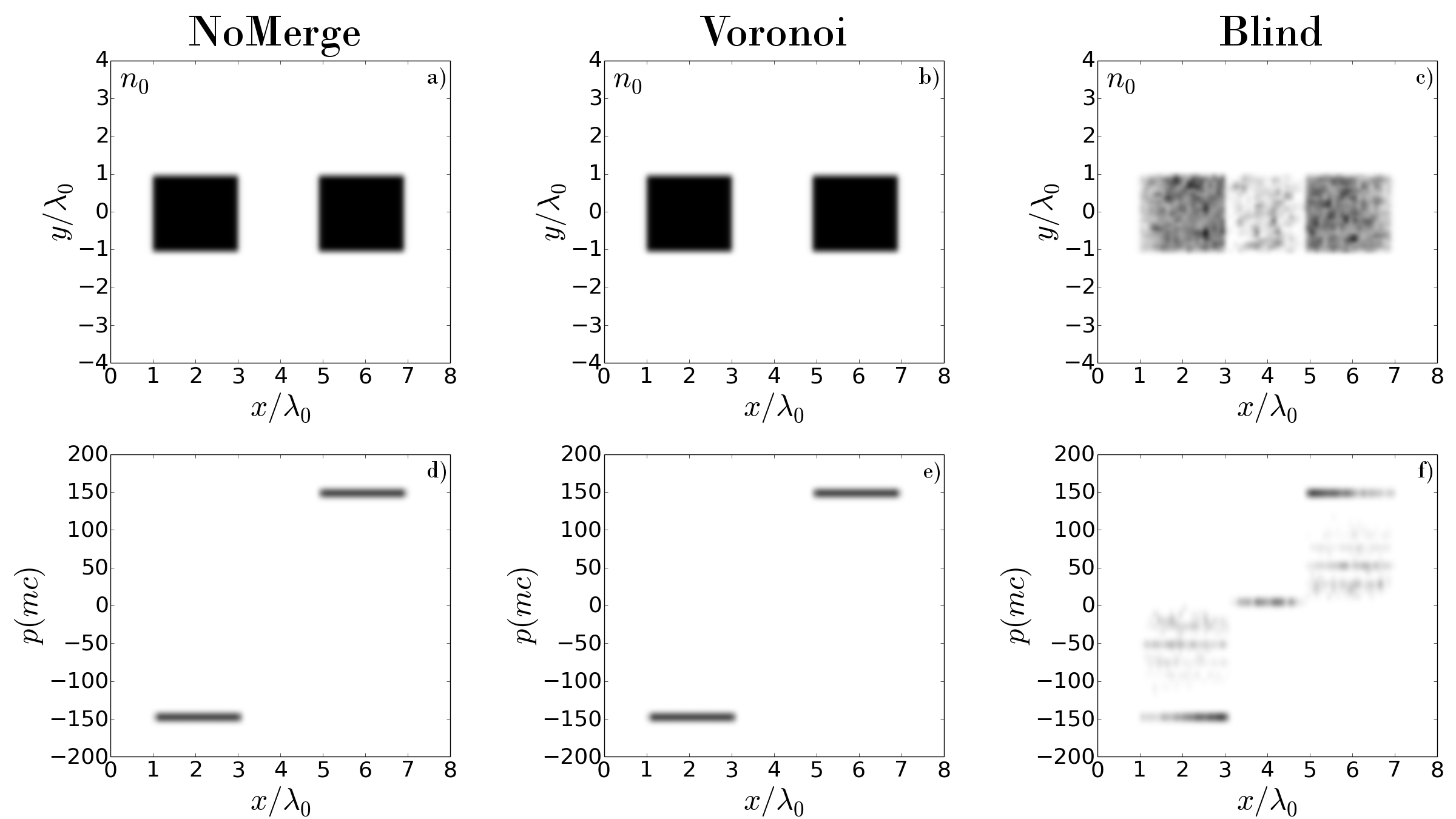}
	\centering
	\caption{The phase space distributions (first row $x/y$, second row $x/p_x$) at the end of the counter-propagating blocks simulation. The blind method leaves behind many particles that have zero momentum. Meanwhile, the Voronoi PMA reproduces the result obtained with no merging.}
	\label{fig:block_dist_last_frame}
\end{figure}

We look at the number of PIC particles appearing in the simulation (see Fig. \ref{fig:block_N_spectrum_w_legend}a). Starting with $12000$ particles, the blind method merges into $4200$ particles at the end of the simulation, while the Voronoi PMA finishes the task with approximately $3800$ particles. The numbers of particles produced by two methods are approximately equivalent. Fig. \ref{fig:block_dist_last_frame} shows the phase space distributions at the end of the simulation and figs. \ref{fig:block_N_spectrum_w_legend} (b,c, and d) show the histogram. For the blind method, we see that after the blocks have passed through each other, there are many particles left behind between the two blocks. The momentum space plot and the histogram shows that these particles have zero momentum. The blind method also produces many particles with momenta not equal to the original magnitude ($150 mc$). As a consequence, the particle distributions are smeared and the conservation of energy is violated. Conversely, the Voronoi PMA accurately preserves the phase space distributions, returning the same result as for the case with no merging. For this test, we see that despite the fact that it finishes the simulation with fewer particles than the blind method, the Voronoi PMA accurately preserves the particle distributions, while the blind method does not.

\subsection{Two-stream instability}
The two-stream instability consists of two identical particle beams streaming through each other. These beams propagate in the opposite directions and a small perturbation in the charge density can change the electric field, which in turn causes further perturbation in the density distributions. This type of simulation makes an illustrative example of how the algorithm manage merging particles in a dynamic evolution of the phase space. The configuration for the two-stream instability is listed in table \ref{table:sim_tsi_config}. At the beginning of the simulation, we create two electron beams with the same initial Lorentz factor $\gamma = 1$ but opposite propagation directions. Each beam has $16 \times 10^4$ particles and is neutralised by the background charge density. Purposefully, the merging algorithms are only enabled after time $t = 5 \lambda_0/c$, when the instability can be visibly observed. The merging fraction for the blind method is chosen to be $\alpha = 0.965$, such that we can have a fair comparison between two algorithms.
\begin{center}
\begin{longtable}{ p{0.5\textwidth} | p{0.3\textwidth} } 
\hline \toprule
	{Wavelength} & {$\lambda_0 = 800 \; \text{nm}$} \\
	{Simulation box}& {$3.2 \lambda_0 \times 1.0 \lambda_0$ } \\
	{Grid steps} & {$0.01 \lambda_0 \times 0.1 \lambda_0$} \\
	{Time step} & {$\Delta t = 0.005 \lambda_0 / c$} \\
	{Electron initial Lorentz factor} & {$\gamma = 1.0$} \\
	{Number of CPUs} & {$8 \times 1$}\\
	{Merging period} & {$50\Delta t$} \\
	{Merging start} & {$5 \lambda_0 / c$} \\
	{The minimum particle number per cell (for Voronoi PMA)} & {$200$} \\
	{Tolerances (for Voronoi PMA)} & {$T_{\mathbf{X}} = 0.8$ and $T_{\mathbf{P}} = 0.15$} \\
	{Merging fraction (for the blind method)} & {$\alpha = 0.965$} \\
		\bottomrule \hline
\caption{The configuration for the two-stream instability simulation.}
\label{table:sim_tsi_config}
\end{longtable}
\end{center}
\begin{figure}[ht!]
	\includegraphics[width=120mm]{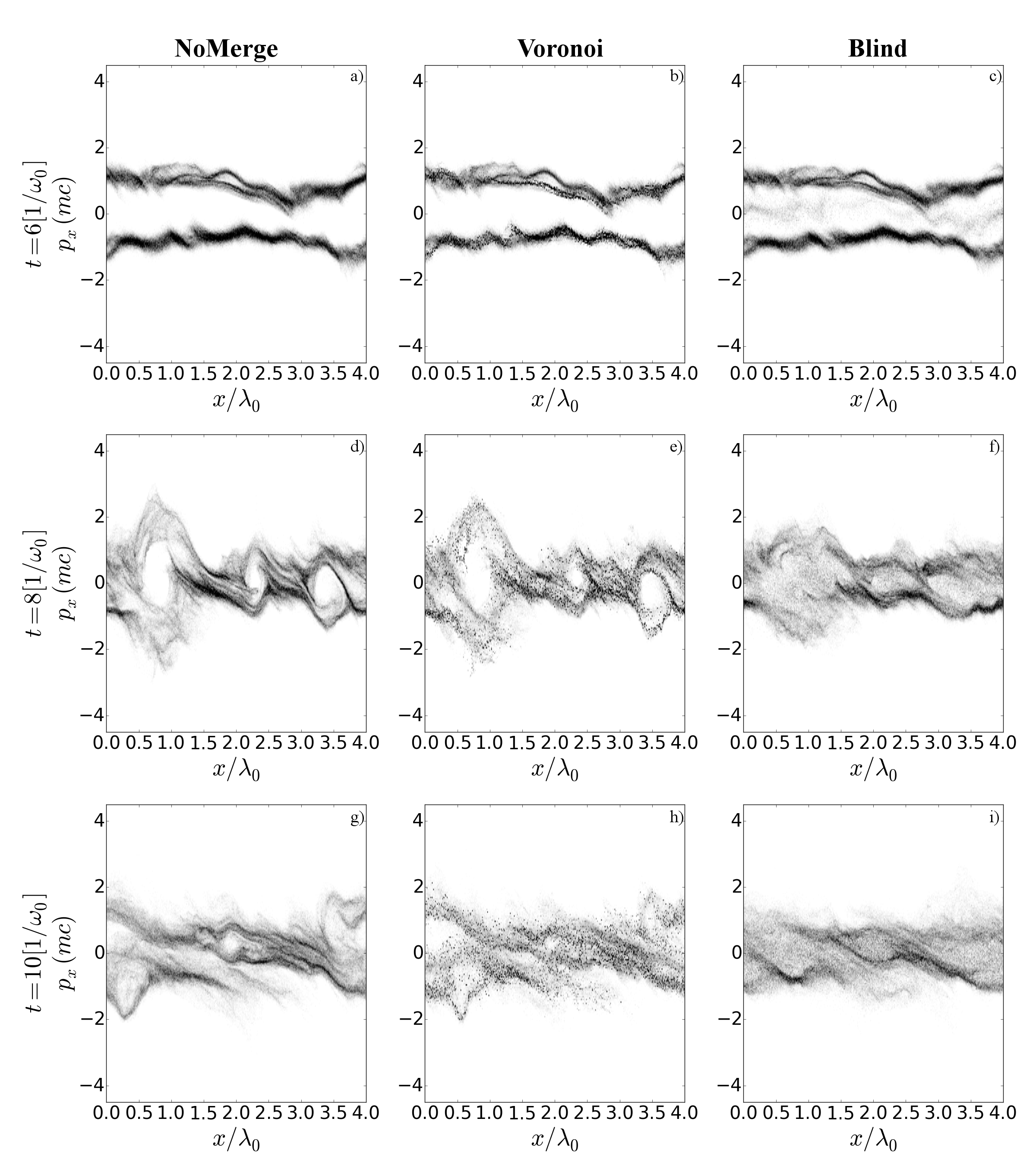}
	\centering
	\caption{The phase space distributions ($x/p_x$) for the two-stream instability simulation at different time stamps. The first column shows the original simulation with $32 \times 10^4$ particles. The second and third columns show the simulation with the blind and Voronoi merging method, respectively. While the outcome produced by the blind method looks different, the Voronoi PMA follows the evolution course as in the no-merging case.}
	\label{fig:tsi_phasespace}
\end{figure}
\begin{figure}[ht!]
	\includegraphics[width=120mm]{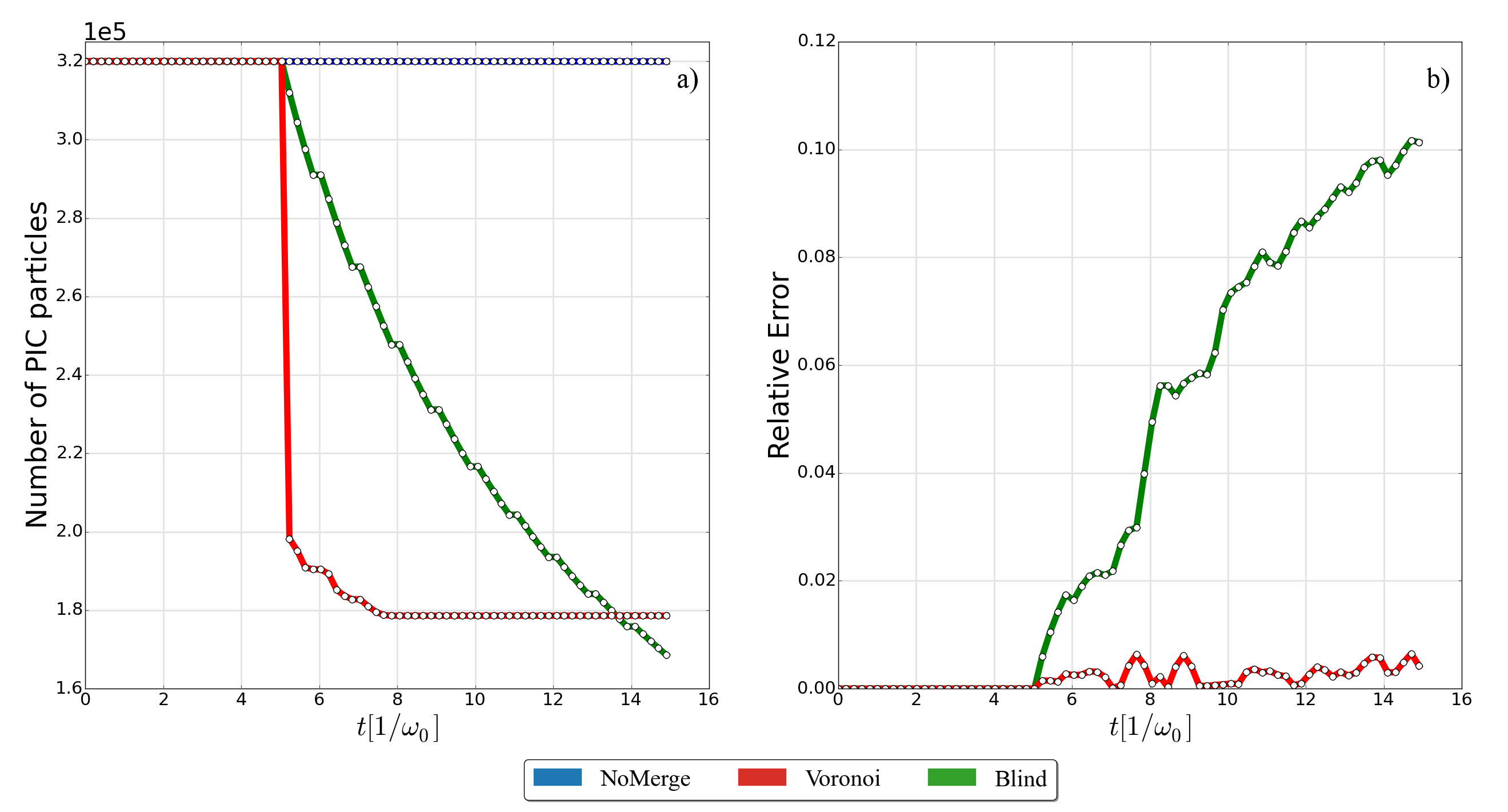}
	\centering
	\caption{The number of PIC particles during the simulation (fig. a) and the relative error in the total energy due to merging events for the two-stream instability simulation. The Voronoi PMA reduces the number of particles from $32 \times 10^4$ to $17.8 \times 10^4$ particles and stops merging from there, since the number of particles per cell is already below the threshold.  The highest relative error in the total energy for the Voronoi PMA is $0.006$, and the blind method $0.1$.}
	\label{fig:tsi_energy}
\end{figure}
The phase space distribution ($x/p_x$) for the two-stream instability is shown in Fig. \ref{fig:tsi_phasespace} at different time stamps. Similarly to the counter-propagating plasma blocks, the blind method (the last column) produces many particles with momenta approximately equal to zero, which do not appear in the original simulation (the first column). This early distortion in the phase space distribution leads to a different instability growth at later time. On the other hand, the Voronoi PMA (the second column) retains the phase space distribution throughout the simulation. Moreover, in contrast to the smooth pictures obtained without merging, the outcomes of the two algorithms appear grainier, since there are lesser particles in the phase space due to merging events. Fig. \ref{fig:tsi_energy} shows the number of electrons and the relative error in the total energy $\delta E = \text{abs}(E_0 - E) / E_0$. Here, $E_0$ is the energy of the system without merging. Observing Fig. \ref{fig:tsi_energy}a, we see that when the merging event is enabled (at $t = 5 \lambda_0 / c$), there is a steep fall in the number of particles for the Voronoi PMA (the red line), falling from $32 \times 10^4$ to $20 \times 10^4$ particles. This abrupt drop is then followed by a short decline to $17.8 \times 10^4$ particles. At around $t = 8 \lambda_0 / c$, there is almost no merging event till the end of the simulation, since the number of particles per cell is already below the threshold. On the contrary, the blind method (the green line) exhibits a steady decline in the number of particles , reducing to $16.8 \times 10^4$ particles at the end of the simulation. From Fig. \ref{fig:tsi_energy}b, wee see that the total energy relative error is rising up to $0.1$ for the blind method, while the Voronoi PMA reaches a peak at $\delta E = 0.006$ during the simulation.
\subsection{Magnetic Shower}
\subsubsection{Introduction}
Consider an energetic particle propagating through a strong magnetic field. Due to the interaction with the field, the particle will emit hard photons on its course. In turn, these photons interact with the field and will decay into energetic electron-positron pairs. The cascade of particles develops quickly and an exponential growth of the number of particles is usually observed. This phenomena is called the magnetic shower. The occurrence of the magnetic shower requires both an intense field and high particle energies \cite{anguelov_showers_1999} \cite{erber_1966}. This condition is quantified in the quantum parameter $\chi$ \cite{anguelov_showers_1999}, which is defined as
\begin{align}
\chi &= \gamma \frac{B}{B_{S}}.
\label{eq:chi_b}
\end{align}

Here, $\gamma$ is the particle's Lorentz factor, $B$ the magnetic field strength, and the Schwinger field $B_S = 4.41 \times 10^{13} \; \text{G}$. The pair production has sufficient probability to start the cascade process only when $\chi \geq 0.1$ \cite{anguelov_showers_1999}. The probability rates for photon emission and pair production are expressed in intricate expressions (see eq. (2) and (3) in ref. \cite{elkina_qed_cascades_2011} and the description therein). The computation usually requires solving the double integral of the Airy function. Thus, the task involves a significant computational overhead. However, under the assumption that the dimensionless field amplitude $a_0 \gg 1$, the field can be regarded as being constant during the decay processes.  Additionally, if both conditions $\chi \gg B / B_S$ and $B \ll B_S$ are satisfied, we can utilise the theory of quantum processes under a constant cross field given in \cite{nikishov_pair_production_1967} \cite{landau_book4}. According to this theory, the probability rates for the photon emission $W_{em}$ and pair production $W_{pair}$ are
\begin{align}
W_{em} &= \frac{\alpha}{3 \sqrt{3} \pi} \frac{mc^2}{\hbar \gamma} \int_0^1 \frac{5x^2 + 7x + 7}{(1 + x)^3} K_{2/3} \left( \frac{2x}{3\chi} \right) \text{d} x
\end{align}
and
\begin{align}
W_{pair} &= \frac{\alpha}{3 \sqrt{3} \pi} \frac{m^2 c^4}{\hbar \varepsilon} \int_0^1 \frac{9 - x^2}{1 - x^2} K_{2/3} \left( \frac{8}{3 (1 - x^2) \kappa} \right) \text{d} x.
\end{align}
Here, $\alpha$ is the fine structure constant; $K_{2/3}(x)$ is the modified Bessel function of the second kind; $\varepsilon$ is photon's energy and $\kappa$ its quantum parameter. Our numerical model for the cascade process is based on the Monte Carlo method \cite{elkina_qed_cascades_2011} \cite{nerush_pair_production_2011}.

The magnetic shower is an appropriate example since the number of particles can grow exponentially during the simulation and the particles' energies can range from several to hundred MeVs. Thus, it is a good indicator of how a PMA copes with the dynamic development during the simulation while preserving the physical features of the system.
\subsubsection{Simulation}
The simulation configuration for the magnetic shower is listed in table \ref{table:sim_mag_config}. We begin the simulation with $5$ numerical electrons. For an electron with a Lorentz factor $\gamma = 5 \times 10^4$ and a magnetic field $eB / m_e c \omega = 500$, the quantum parameter is $\chi \approx 150 \gg 1$. Here, $e$ is the elementary charge, $m_e$ the electron mass, $c$ the velocity of light, and $\omega = 2\pi c / \lambda_0$. As before, we consider three cases: without merging, with the blind merging method, and with the Voronoi algorithm. As before we deliberately choose the merging fraction $\alpha$ such that the blind method and the Voronoi PMA result in the similar number of particles at the end of the simulation.
\begin{center}
\begin{longtable}{ p{0.5\textwidth} | p{0.3\textwidth} } 
\hline \toprule
	{Wavelength} & {$\lambda_0 = 800 \; \text{nm}$} \\
	{Simulation box}& {$3.2 \lambda_0 \times 3.2 \lambda_0 \times 3.2 \lambda_0$} \\
	{Grid steps} & {$0.04 \lambda_0 \times 0.04 \lambda_0 \times 0.04 \lambda_0$} \\
	{Time step} & {$\Delta t = 0.005 \lambda_0 / c$} \\
	{Magnetic field strength} & {$B = 6.6 \times 10^{10} \; \text{G}$} \\
	{Electron initial Lorentz factor} & {$\gamma = 5 \times 10^4$} \\
	{Number of CPUs} & {$5 \times 5 \times 5$}\\
	{Merging period} & {$2\Delta t$} \\
	{The minimum particle number per cell (for Voronoi PMA)} & {$10$} \\
	{Tolerances (for Voronoi PMA)} & {$T_{\mathbf{X}} = 1.0$ and $T_{\mathbf{P}} = 0.02$} \\
	{Merging fraction (for the blind method)} & {$\alpha = 0.88$} \\
		\bottomrule \hline
\caption{The configuration for the magnetic shower simulation.}
\label{table:sim_mag_config}
\end{longtable}
\end{center}
\begin{figure}[ht!]
	\includegraphics[width=120mm]{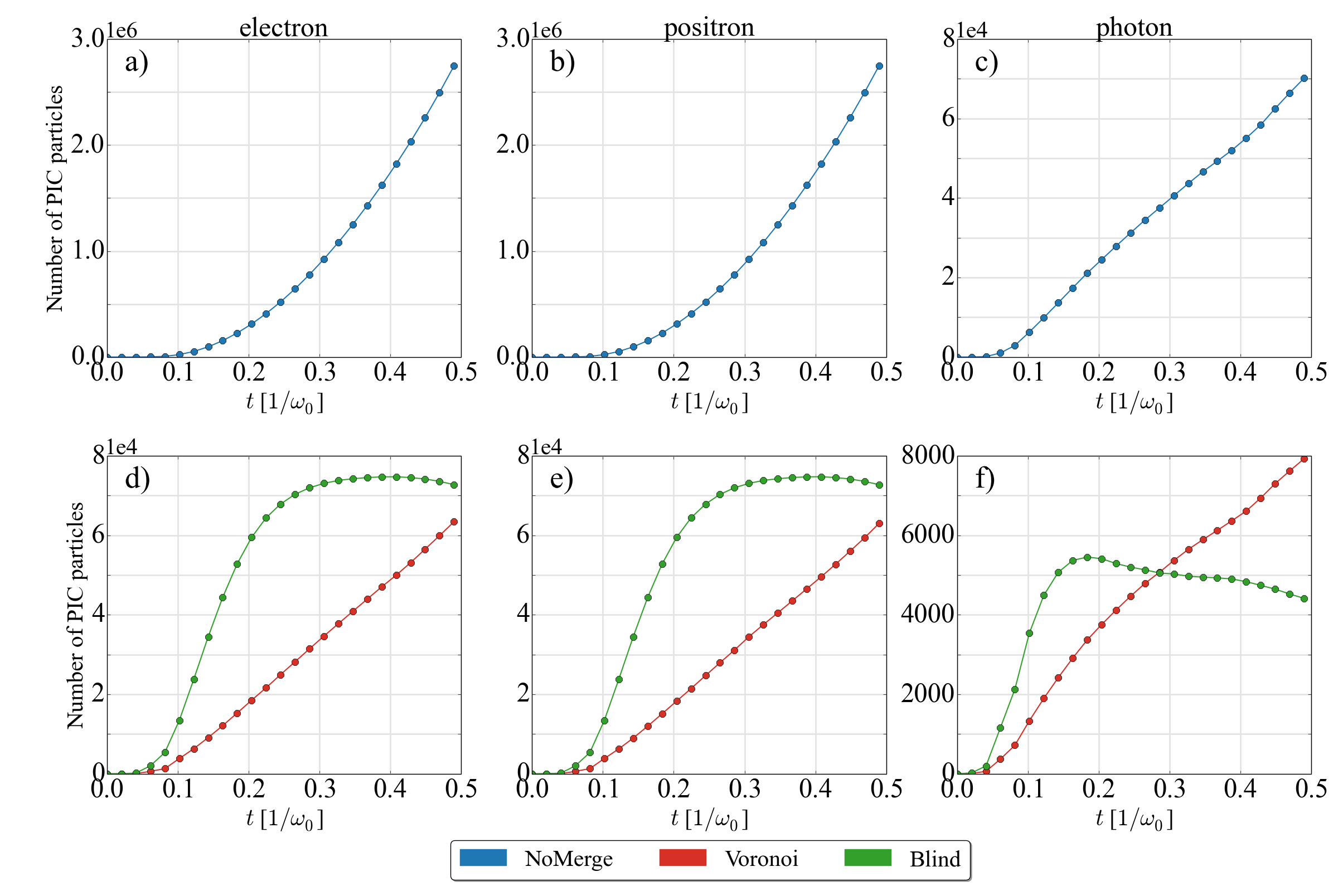}
	\centering
	\caption{The number of particles in the magnetic shower simulation as a function of time for electron, positron, and photon (from left to right). The first row shows the result from the simulation without merging, the second row shows the outcome by using the Voronoi PMA (red) and the blind method (green). The Voronoi PMA reduces the number of electrons (positrons) from $2.8 \times 10^6$ to $6.5 \times 10^4$ particles. }
	\label{fig:total_N_w_legend}
\end{figure}

\begin{figure}[ht!]
	\includegraphics[width=120mm]{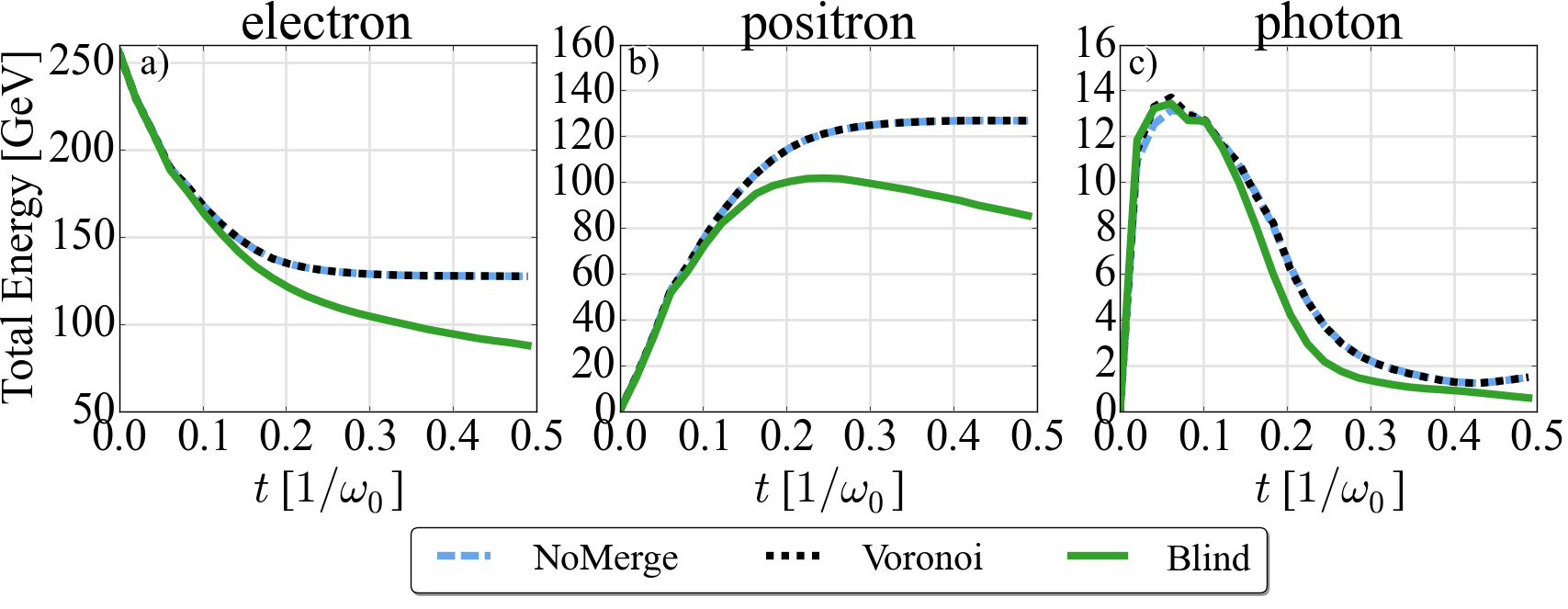}
	\centering
	\caption{The total energy evolution of electron, positron, and photon during the magnetic shower simulation for three merging cases: no-merge case (long dash, light blue); Voronoi (short dash, black), and blind (solid, green). Unlike the blind method, the Voronoi PMA reproduces the results obtained from the original simulation.}
	\label{fig:total_energy_w_legend}
\end{figure}

\begin{figure}[ht!]
	\includegraphics[width=120mm]{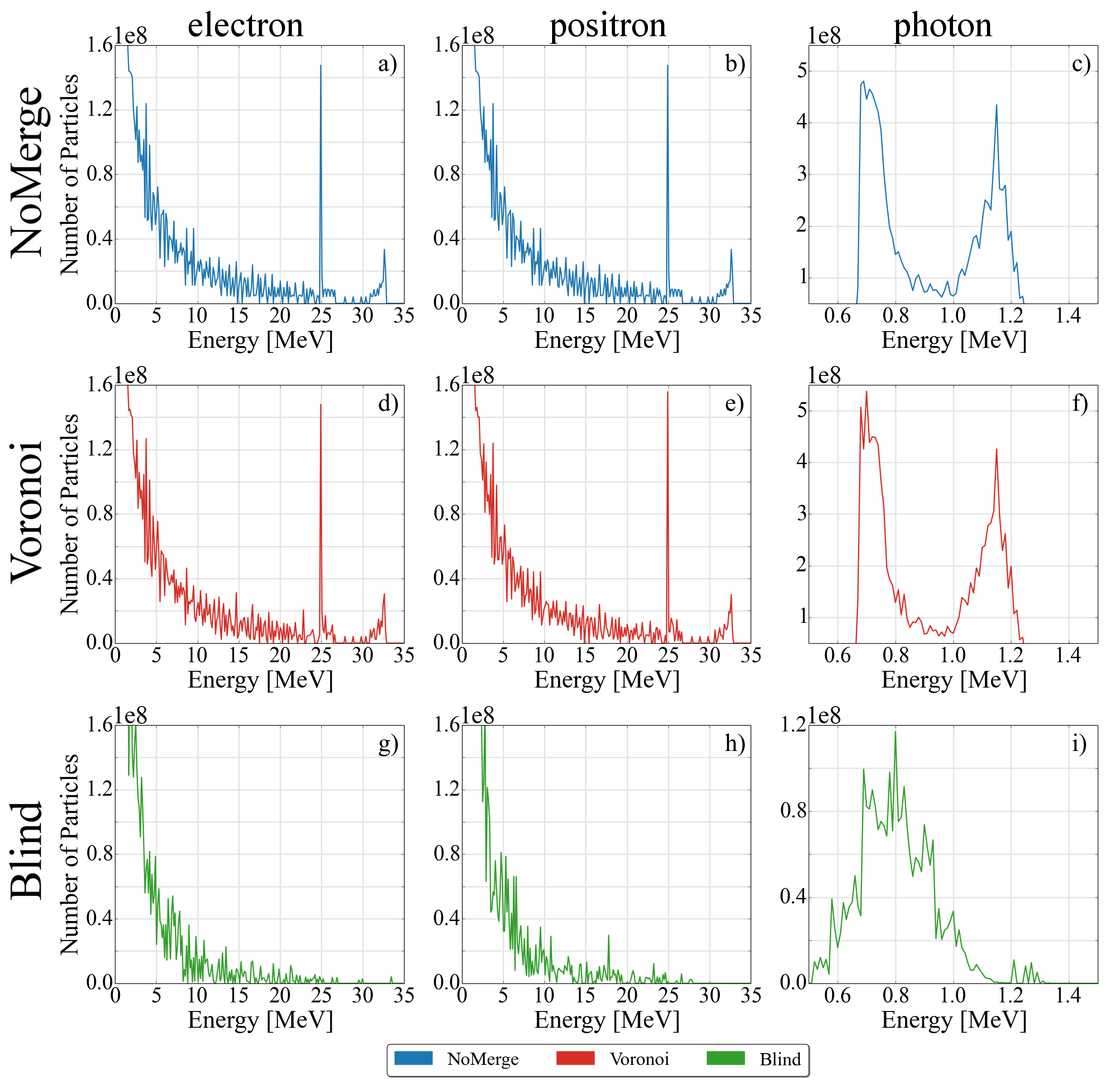}
	\centering
	\caption{The spectra for the electron, positron, and photon species in the magnetic shower simulation at time $t = 98 \Delta t$ for three merging cases: no merge (blue), the Voronoi PMA (green), and the blind method(red). The spectra of particles are accurately reproduced by using the Voronoi PMA. Meanwhile, with the blind method, the distinct peak for electrons and positrons is not observed.}
	\label{fig:spectrum_magnetic_shower}
\end{figure}

The growth in particle number is shown in Fig. \ref{fig:total_N_w_legend}. Without merging (blue), both electron and positron display exponential growth during the simulation. At the end of the simulation, a total number of $2.8 \times 10^6$ particles has been reached for each specie. Meanwhile, the photon specie grows from $0$ to $7 \times 10^4$ particles at the last frame. The blind method (green) results in $1.45 \times 10^5$ electrons and posittrons, $4000$ photons. The Voronoi PMA (red) produces in total $1.35 \times 10^5$ electrons and positrons, and $8000$ photons. That is, the number of particles in the box is reduced approximately $40$ times by both methods. In order to verify the validity of the simulation, we look at the total energy and the spectra of the particles. Figs. \ref{fig:total_energy_w_legend} and \ref{fig:spectrum_magnetic_shower} illustrate the evolution of the particle energies and their spectra at the end of the simulation. For the blind method (solid, green line in Fig. \ref{fig:total_energy_w_legend}), we see a gradual decrease in the total energy of electrons and positrons around the point when the photon energy is reaching its peak. This strongly affects the spectrum of every specie in the simulation box (see Figs. \ref{fig:spectrum_magnetic_shower} g, h, i): the distinct peak electrons and positrons is not observed. On the other hand, with a careful approach the Voronoi PMA (short dash, black line) overlaps the case with no merging (long dash, light blue )in Fig. \ref{fig:total_energy_w_legend}, showing that it preserves the physical behaviour in the total energy, with the decrease in electron energy, increase in positron energy, and the sharp rise followed by a decrease in photon energy. Moreover, the Voronoi PMA accurately reproduces the spectra obtained with no merging (see Figs. \ref{fig:spectrum_magnetic_shower} d, e, and f). Originally, the simulation with no merging takes approximately $2$ hours ($7265$ seconds). With the same settings, the Voronoi PMA completes roughly in $20$ minutes ($1172$ seconds) and the blind method takes about $24$ minutes ($1440$ seconds).

\begin{figure}[ht!]
	\includegraphics[width=120mm]{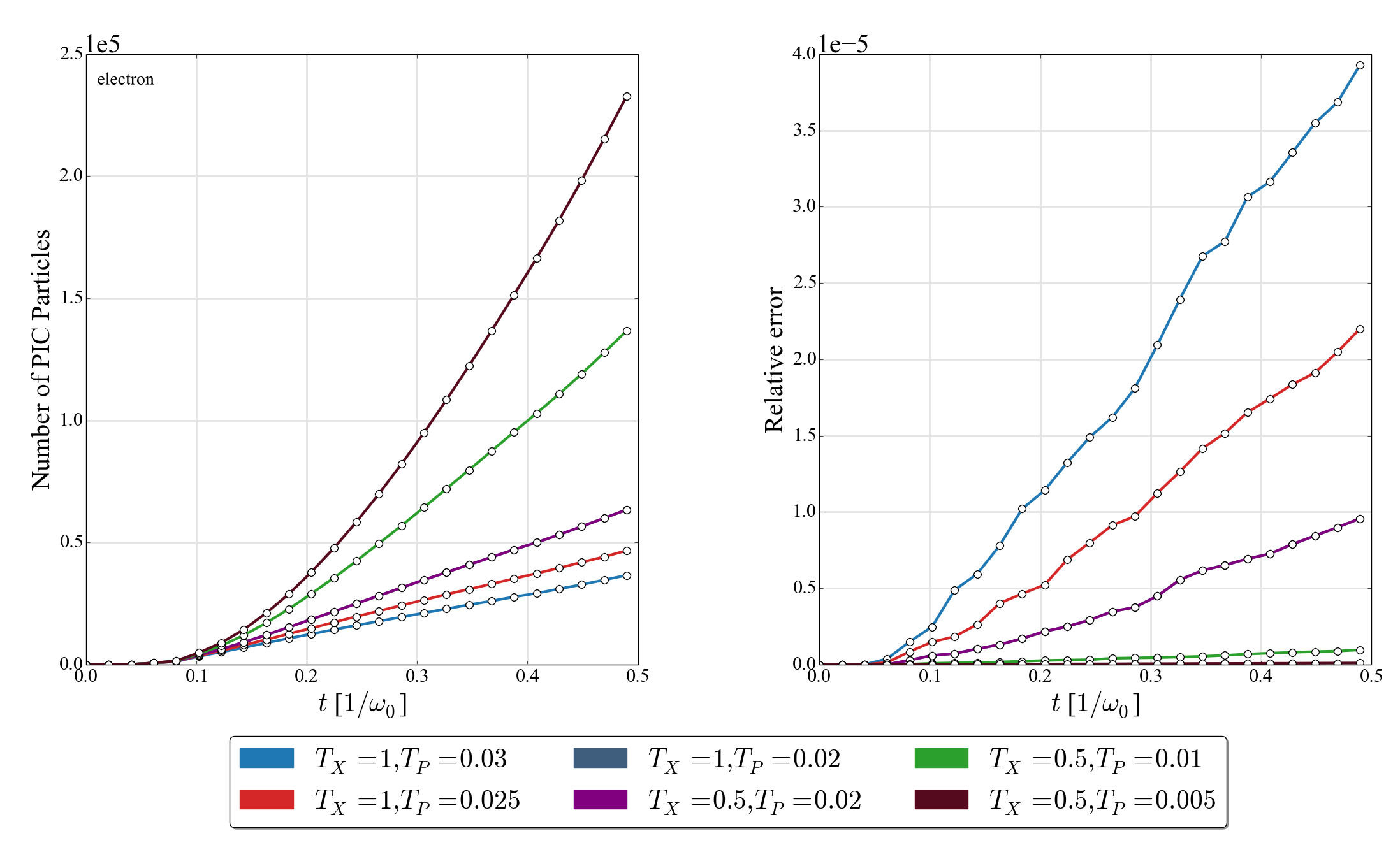}
	\centering
	\caption{The number of electron and the relative error in the total energy due to merging with various tolerances $[T_\mathbf{X}, T_\mathbf{P}]$ settings for the magnetic shower simulation. With relaxed tolerances, the growth of particle number becomes linear but the error also accumulates faster. When stricter tolerances are used, the growth resumes the exponential behaviour while the error develops with a slower rate. We also observe that, the purple line ($T_\mathbf{X} = 0.5$ and $T_\mathbf{P} = 0.02$) completely overlaps the dark blue line ($T_\mathbf{X} = 1.0$ and $T_\mathbf{P} = 0.02$), showing that the algorithm is always more sensitive towards the momentum space.}
	\label{fig:voronoi_compare_w_legend}
\end{figure}

\begin{figure}[ht!]
	\includegraphics[width=120mm]{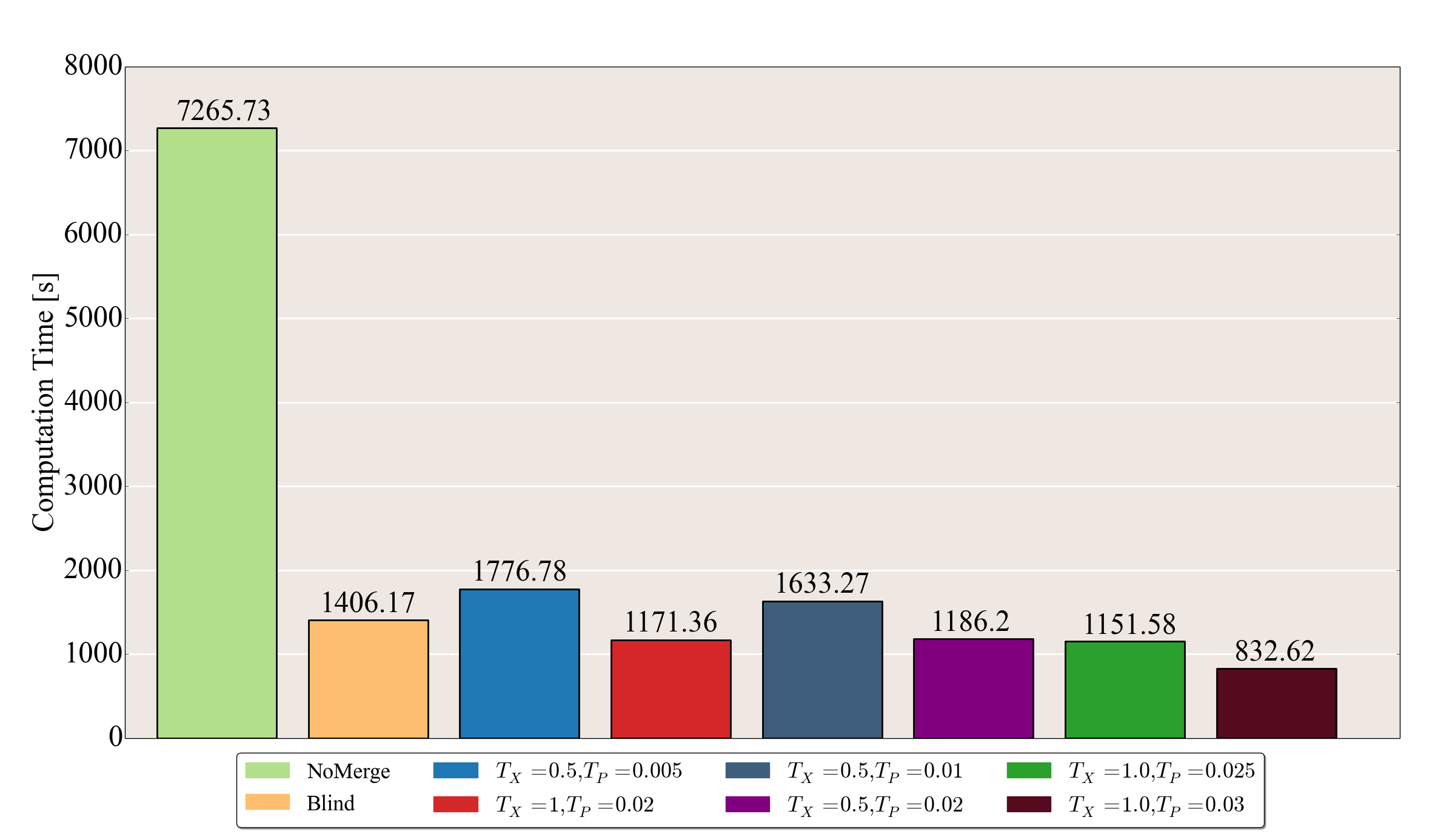}
	\centering
	\caption{The total computation time for the magnetic shower simulation. The last six columns show the simulations with the Voronoi PMA for various tolerances $[T_\mathbf{X}, T_\mathbf{P}]$ settings.}
	\label{fig:comp_time}
\end{figure}

Finally, we perform a parameter scan on the tolerances $T_\mathbf{X}$ and $T_\mathbf{P}$ in order to observe the growth of particles and the accumulation of error due to merging. Fig. \ref{fig:voronoi_compare_w_legend} shows the number of electrons and the relative error $\delta E = (E_0 - E) / E_0$ during the simulation and Fig. \ref{fig:comp_time} displays the total computation time with various tolerance settings. Here, $E_0$ is the energy of the system without merging. The most accurate simulation is achieved with $T_\mathbf{X} = 0.5$ and $T_\mathbf{P} = 0.005$. With this setting, the simulation takes roughly $40$ minutes to complete and the total energy loss is around $0.05$ MeV ($\delta E \approx 1 \times 10^{-7})$. We observe that the growth is also exponential and the number of electrons has reached $2.4 \times 10^5$ particles at the end of the simulation. When we loosen the tolerances, more particles are merged together. As a result, the growth rate becomes more linear but the energy loss develops speedily. In our test, the extreme case with $T_\mathbf{X} = 1.0$ and $T_\mathbf{P} = 0.03$ produces $7.7 \times 10^4$ electrons and positrons, $5 \times 10^3$ photons, and takes $14$ minutes to finish. However, in this case, it accumulates $20$ MeV total energy loss ($\delta E \approx 3.9 \times 10^{-5})$. Although the loss is extremely small, we notice the double in magnitude just by increasing from $T_\mathbf{P} = 0.025$ to $T_\mathbf{P} = 0.03$. We also observe that, the purple line ($T_\mathbf{X} = 0.5$ and $T_\mathbf{P} = 0.02$) completely overlaps the dark blue line ($T_\mathbf{X} = 1.0$ and $T_\mathbf{P} = 0.02$), showing that the tolerance $T_\mathbf{P}$ is more sensitive than $T_\mathbf{X}$. Since, in a given cell, the particle momenta may vary significantly, an accurate simulation requires small $T_\mathbf{P}$. We recommend $T_\mathbf{P} = 0.05$ and $T_\mathbf{X} = 1.0$ as a threshold for this type of simulation.
\section{Summary} \label{s_summary}
In this paper, we present the Voronoi particle merging algorithm for PIC codes. The phase space of a simulation cell is partitioned, as in the Voronoi diagram, into smaller subsets, which only consist of particles that are close to each other. The quality of a merging event is ensured by two user inputs, the tolerances on position $T_\mathbf{X}$ and momentum $T_\mathbf{P}$. The tolerances act as the balance between the speed-up and the accuracy of the simulation. Stricter tolerances mean smaller error but without much in the speed-up. On the other hand, relaxed tolerances result in more merged particles and thus the computation time decreases but the error will accumulate faster. Making a right combination for the tolerance pair for a certain simulation requires prior knowledge of particles' behaviour. If a simulation involves particles which spread out in a large range in the momentum space, we suggest keeping the $T_\mathbf{P}$ lower than $0.02$. Otherwise, this value can be relaxed. On the other hand, since it relates to particles' relative position in a cell, $T_\mathbf{X}$ can be chosen up to $1.0$.

We have tested the performance of our algorithm with three tests: the counter-propagating plasma blocks, the two-stream instability and magnetic shower simulations. In all cases, we observe that the conservation of momentum is perfectly held and the conservation of energy is maintained extremely well, with only small margin of error. The two-stream instability shows that the Voronoi PMA preserves the phase space evolution and the total energy error in this case is of the order of $10^{-3}$. In the magnetic shower simulation, the total energy error is of the order of $10^{-5}$ with a speed-up by a factor of $6$ and the spectra of particles are also comparable very well to those obtained with no merging.

The authors would like to thank Dr. John Farmer and Axel H\"ubl for many fruitful discussions.

This work has been supported by the Deutsche Forschungsgemeinschaft via GRK 1203 and SFB TR 18, by BMBF (Germany), and by EU FP7 project EUCARD-2.

\section*{References}

\bibliography{cavalier}

\end{document}